\documentclass[twocolumn,showpacs,preprintnumbers,amsmath,amssymb]{revtex4-1} 
\usepackage{color}

\usepackage[utf8]{inputenc}
\usepackage[colorlinks,hyperindex]{hyperref}  
\hypersetup
{
  colorlinks,%
  citecolor=blue,%
  linkcolor=blue,%
  urlcolor=blue,%
}

\def\setR{\mathbb{R}}

\def\nlb{\nolinebreak}
\DeclareMathOperator{\sgn}{sgn}

\newcommand{\norm}[1]{\parallel\!#1\!\parallel}

\newcommand{\sss}[1]{\scriptscriptstyle #1}

\begin{document}

\title{FLRW spaces as submanifolds of $\setR^6$: restriction to the Klein-Gordon operator}

\author{J. P. Arias Zapata$^1$}
\author{A.~Belokogne$^2$}
\author{E.~Huguet$^1$}
\author{J.~Queva$^3$}
\author{J.~Renaud$^4$}
\affiliation{$1$ - Universit\'e Paris Diderot-Paris 7, APC-Astroparticule et Cosmologie (UMR-CNRS 7164), 
Batiment Condorcet, 10 rue Alice Domon et L\'eonie Duquet, F-75205 Paris Cedex 13, France.\\
$2$ - Universit\'e de Corse, APC-Astroparticule et Cosmologie (UMR-CNRS 7164), 
Batiment Condorcet, 10 rue Alice Domon et L\'eonie Duquet, F-75205 Paris Cedex 13, France.\\
$3$ - Equipe Physique Th\'eorique, Projet COMPA, SPE (UMR 6134), 
Universit\'e de Corse, BP52, F-20250,  Corte, France.\\
$4$ - Universit\'e Cergy-Pontoise, APC-Astroparticule et Cosmologie (UMR-CNRS 7164), 
Batiment Condorcet, 10 rue Alice Domon et L\'eonie Duquet, F-75205 Paris Cedex 13, France.
} 
\email{jeanphilippe.ariaszapata@gmail.com, \\belokogne.andrei@gmail.com, \\huguet@apc.univ-paris7.fr, \\queva@univ-corse.fr,\\
jacques.renaud@apc.univ-paris7.fr}

\date{\today}


\begin{abstract}
The FLRW spacetimes can be realized as submanifolds of $\setR^6$. In this paper we relate the 
Laplace-Beltrami operator for an homogeneous scalar field $\phi$ of $\setR^6$ to its explicit restriction on FLRW spacetimes. We then 
make the link between the homogeneous solutions of the equation $\square_6 \phi = 0$ in $\setR^6$ and those of the Klein-Gordon 
equation  $(\square_{f} - \xi R^f  + m^2)\phi^f=0 $ for the free field $\phi^f$ in the FLRW spacetime.
We obtain 
as a byproduct a formula for the Ricci scalar of the FRLW spacetime in terms of the function $f$ defining this spacetime in  $\setR^6$.   

\end{abstract}

\maketitle

\section{Introduction}\label{Sec-Introduction}

The aim of the present article is to restrict the scalar Laplace-Beltrami operator defined on $\setR^6$, endowed with the metric 
$\eta~=~\mbox{diag}(+,-,-,-,-,+)$,  to (locally) conformally flat four dimensional Lorentzian manifolds $X_f$. These manifolds are 
obtained as the intersection
between the null cone $\mathcal{C}$ and the submanifolds $P_f$ of $\setR^6$, defined through functions $f$ homogeneous of degree one. 
The set of the manifolds $X_f$ contains in particular the Friedmann-Lema\^{\i}tre-Robertson-Walker (FRLW) spacetimes. The definition of a 
manifold $X_f$,  as the intersection of two 
hypersurfaces, leads to the definition of two natural vector fields: schematically the normals to the hypersurfaces. From a geometrical 
viewpoint these two vector fields operate on scalar fields: they are constraints associated to the manifold $X_f$. These two constraints 
can be used to obtain the restriction of the scalar Laplace-Beltrami operator to the manifold $X_f$. Our strategy to obtain that restriction
is to build an orthonormal basis (a sextrad) in the neighborhood of $X_f$ which contains the constraints and to use a geometrical identity
to split the Laplace-Beltrami operator, which in this case identifies with the Laplace-de~Rham operator up to a sign, into the Laplace-Beltrami 
operator on $X_f$ and remaining terms related to the constraints. 

On a physical side, the restriction from $\setR^6$ to the submanifold $X_f$ maps a scalar field solution of the free ``massless'' 
Klein-Gordon equation in $\setR^6$ to a solution of the corresponding restricted equation
on $X_f$. Now, it appears that sufficient freedom
is left in the constraints (which apply on the scalar field) to choose them in such a way that 
the restricted field on $X_f$ satisfies the Klein-Gordon-like equation (possibly massive and coupled to the curvature): 
$(\square_{f} - \xi R^f  + m^2)\phi^f=0,$  on $X_f$. That is in particular on FLRW spacetimes. Note that, the term $m^2$ (when present) is 
only a coefficient in a curved space, we nevertheless keep the name of Klein-Gordon  (instead of Klein-Gordon-like) 
for short, in the remaining of the text.

At the origin we developed part of the geometric formalism we use here to study the properties of conformally invariant fields on (A)dS 
spacetimes and how these properties relate to their Minkowskian counterparts if any \cite{Huguet:2006fe}. One of the main advantage of 
this geometric setup is
that it provides us with a (local) continuous mapping of mathematical objects (functions, operators, etc.) between spaces. Indeed, this 
formalism applies not only to maximally symmetric spaces, but to all locally conformally flat spaces \cite{Huguet:2013dia, Huguet:2013tv}. 
As a consequence, 
non-maximally symmetric FLRW spacetimes are concerned, which makes contact with cosmology (FLRW spacetimes,  
are currently favored, on large scale, by recent obervations \cite{Planck:2016}, with possible deviation from spatial 
flatness \cite{Sola:2016jky}). 
In that context it is of interest to consider, beside 
conformally invariant massless field, the possibility for non-conformally coupled fields and massive fields.

Finally, it should be emphasized that, if the realization of some spacetimes as 
submanifolds of some ambient spaces is known for some time (see for instance \cite{Bayen:1984dt}, \cite{Binegar:1982fv} for conformal 
field in $\setR^6$ and more generally
references in \cite{Gulamov:2011ux}), the formalism we use here differs from 
these methods in a notable way. 
Its specificity lies in the realization of the four dimensional space $X_f$ as 
an intersection between the null cone of $\setR^6$ and the submanifold $P_f$ 
defined by the function $f$. This submanifold $P_f$ can be considered as an 
embedding space which is adapted for each spacetime $X_f$. This differs, from 
this point of view, with so-called ambient space formalism (see for instance 
\cite{Gazeau:1999mi, Penedones:2016voo}) in which the embedding space is fixed.

The paper is organized as follows. The restriction of the scalar Laplace-Beltrami operator from $\setR^6$ to $X_f$  
is done in Section \ref{SEC-RestrictioToXf}. As a byproduct one obtains a formula  relating the Ricci scalar $R^f$ on the spacetime 
and the function $f$ defining that spacetime in $\setR^6$. The Section \ref{SEC-KGForm} is devoted to the relation with the Klein-Gordon equation
on the spacetime $X_f$. We conclude in Section \ref{SEC-Conclusion}. Some topics and additional materials
are collected in appendices~: appendix \ref{App-DefAndForm} details the conventions and notations used in this work, a somewhat technical proof
used for the restriction of the Laplace-Beltrami operator is given in appendix \ref{App-Proof-deld6-deld4}, an explicit realization to general
FLRW spacetimes is given in appendix \ref{App-ConstrainFLRW}, the direct application of our results to the maximally symmetric spaces is 
tackled in appendix \ref{App-ConstraintMaxSym}.

Our conventions for indexes are: $\alpha, \beta, \gamma,\ldots = 0,\ldots, 5$, and $\mu,\nu,\rho,\ldots =  0, 1, 2, 3$.
In addition, except otherwise stated, a superscript (or subscript) $f$ indicates a quantity on the $X_f$ space (see Section \ref{SEC-RestrictioToXf} 
for the definition). The space $\setR^6$ is provided with the natural orthogonal coordinates $\{y^\alpha\}$ and the metric  
$\eta~=~\mbox{diag}(+,-,-,-,-,+)$ which is left invariant under the group O$(2,4)$ and its connected subgroup SO$_0(2,4)$. 
Our notations for differential geometry follow those of Fecko \cite{Fecko:2006}. Certain definitions and formulas particularly 
used in this work are gathered in appendix \ref{App-DefAndForm}.

\section{The restriction of $\square_6$ to the spacetime $X_f$}\label{SEC-RestrictioToXf}
Let us first recall the geometrical framework we use: a four dimensional  manifold $X_f$ is 
obtained as the intersection 
of the five dimensional null 
cone of $\setR^6$: $\mathcal{C}:=\{y\in \setR^6 : C(y) = 0\}$,  where $C(y):=y^\alpha y_\alpha$, 
and the hyper-surface  $P_f:= \{y\in \setR^6 : f(y) = 1\}$, in which $f$ 
is homogeneous of degree one. 
Locally conformally flat spaces, including FLRW, de Sitter and anti-de Sitter spaces,
 can be obtained by suitable choices of the 
homogeneous function $f$. Details about this construction may be found in \cite{Huguet:2013tv}.

In this section we explicitly restrict the free 
scalar equation $\square_6 \phi = 0$ to a four dimensional space $X_f$. 
This is obtained thanks to a geometrical method which refers to Cartan's moving frame (``rep\`ere mobile'') theory.
Indeed, instead of the common procedure which consists in using a specific coordinate system including constraints 
(for instance $\{C(y), y^\mu/f(y), f(y)\}$), we relate constraints to vectors (differential operators) on
which we build the moving frame.
As a consequence, no coordinate system other than the natural Cartesian system of $\setR^6$: $\{y^\alpha\}$ is needed.  

Beside the restriction of the  scalar operator $\square_6$,  we obtain, as a byproduct,
a formula which allows us to compute the Ricci scalar of the space $X_f$ from the function $f$ which defines it as a 
submanifold embedded in $\setR^6$, namely
\begin{equation*}
\frac{R^f}{6} =  (\square_6f)^f - 2 F_f^2,
\end{equation*}
where $F=\widetilde{df}=\partial^\alpha f\,\partial_\alpha$ is the gradient vector of $f$.

\subsection{A general expression for $(\square_6\phi)^f$}\label{closed formula}

The basic idea in order to restrict the scalar operator $\square_6 = -\delta d$ to the submanifold $X_f$ is to use the formula 
(\cite{Fecko:2006}, \S 5.8.11)
\begin{equation}\label{EQ-BoxSpliting}
j^a i_a = p,
\end{equation}
which applies to any p-form in an orthonormal basis $\{e_a\}$, to split the operator
into two parts:
\begin{equation*}
\square_6 \phi = -\delta d\phi = -\delta j^\mu i_\mu d\phi - \delta(j^4 i_4 + j^5 i_5) d\phi,  
\end{equation*}
in such a way that the first term in the r.h.s. 
can be restricted to be the Laplace-Beltrami operator of $X_f$ acting on the restricted field $\phi^f$ and the second term can be
expressed in terms of constraints related to those defining the geometry ($f$ and $C$) and intrinsic quantities of $X_f$. 
In order to complete that scheme, we need a basis $\{e_a\}$ in which we perform 
the splitting in a neighborhood of $X_f$ and to translate the constraints 
defining $X_f$ in operators acting on the scalar field. 

Let us begin by considering the constraints on the field. Since they account for the geometry, that is to the surfaces defined by 
$f$ and $C$ whose intersection is the space $X_f$, a natural choice is to use the two gradient vectors  
\begin{equation}\label{EQ-defFandD}
F := \widetilde{df}=\partial^\alpha f\,\partial_\alpha ,~\mathrm{and}~D:= \frac{1}{2} \widetilde{dC} = y^\alpha \partial_\alpha,
\end{equation}
as constraint operators on the field. Note that $D = y^\alpha \partial_\alpha$ is the generator of dilations in $\setR^6$. 
The commutators 
\begin{equation*}
[\square_6, D] = 2 \square_6,~~~[F, D] =  F,
\end{equation*}
show us that, since $\square_6 \phi = 0$, only $F\phi$ has to be given. The operator $D$ can be left unspecified, which means that 
the function $\phi$ can be choosen homogeneous of some unspecified weight $r$, namely $D\phi = r \phi$.

We now build the orthonormal basis. Since it must be defined in a neighborhood of $X_f$ let us define the space 
\begin{equation}\label{EQ-Def-Mpq}
M_{\lambda,q}=\{y:\; f(y)=\lambda,\, C(y)=q,\; \lambda,q \in \setR, \lambda > 0\}.
\end{equation}
We then have $X_f = M_{1,0}$ and since $X_f$ is a pseudo-Riemannian manifold so is $M_{\lambda,q}$ at least in a neighborhood of $X_f$.
Let us choose for each $M_{\lambda,q}$ an orthogonal bases field of T$M_{\lambda,q}$: $\{\partial_\mu\}$. A vector $\partial_\mu$ is
tangent to the surfaces defined by $f(y)=\lambda$ and to the surfaces defined by $C(y)=q$ it is thus orthogonal to $F$ and $D$. 
We can then build an orthonormal basis $\{e_\alpha\}$ from the six vectors $\{\partial_\mu, F, D\}$. 

We consider two separate cases depending on the nature of $F$. Let us first assume $F^2\neq0$, and set
\begin{equation}\label{EQ-eFe_muFneq0}
e_\mu := \frac{\partial_\mu}{\vert\partial_\mu\vert},~~e_{\sss 5} := \frac{F}{\vert F \vert}.
\end{equation}
Then $e_4$ is obtained  as a normed linear combination of $D$ and $F$,  we find
\begin{equation}\label{EQ-e4-F2neq0}
e_4 =\frac{1}{\sqrt{\Delta}}\left( f e_5-\vert F\vert D\right) ,
\end{equation}
with $\Delta := f^2-F^2 D^2 $. Note that, these expressions apply whatever the sign of $F^2$.

When $F^2=0$ we redefine $e_5$ as
\begin{equation}\label{EQ-e5 F=0}
e_{\sss 5} := \frac{F+D}{\vert F + D\vert}.
\end{equation}
Then, we can complete the two orthonormal vectors $\{e_\mu, e_5\}$ and obtain the orthonormal basis $\{e_\alpha\}$ with
\begin{equation}\label{EQ-e4-F2nq0}
e_4 =\frac{1}{\sqrt{\Delta_0}}\left( (1 + \frac{D^2}{f}) F -  D\right),
\end{equation}
where $\Delta_0 := \vert F + D\vert = 2f + D^2$.

Having determined an orthonormal basis we can now split $\square_6 \phi$ in that basis using (\ref{EQ-BoxSpliting}) making the 
constraints apparent. As mentioned before, 
the term $-\delta j^\mu i_\mu d\phi$ reduces to $\square_f$ on $X_f$, the proof of this statement is given in Appendix \ref{App-Proof-deld6-deld4}. 
The calculation of the second term of the r.h.s. of (\ref{EQ-BoxSpliting})~: $\delta(j^4 i_4 + j^5 i_5) d\phi$, follows the same lines
in the three cases $F^2>0$,  $F^2<0$ or $F^2=0$ and leads to the same final formula. Here, we perform the 
calculation for $F^2>0$. We first express the operators $i$ and $j$ as   
\begin{align*}
i_5&=\frac{1}{|F|}i_{\sss F},\\
i_4&=\frac{f}{\sqrt{\Delta}{|F|}}i_{\sss F}
-\frac{|F|}{\sqrt{\Delta}}i_{\sss D},\\
j^5&=\frac{1}{|F|}j_{\sss F},\\
j^4&=-j_4= -\frac{f}{{\sqrt{\Delta}}|F|}j_{\sss F}
+\frac{|F|}{\sqrt{\Delta}}j_{\sss D}.
\end{align*}
Then, one obtains
\begin{equation*}
\begin{split}
&\delta (j^5i_5+j^4i_4)d\phi=\\ 
&\delta\left(-\frac{D^2}{\Delta}j_{\sss F} F+
\frac{f}{\Delta}j_{\sss F} D+\frac{f}{\Delta}j_{\sss D}F-\frac{F^2}{\Delta}j_{\sss D} D\right)\phi.
\end{split}
\end{equation*}
This expression can be explicitly calculated in the natural Cartesian coordinate system of $\setR^6$ in which the codifferential
operator applied on a one-form $\omega$ is minus the divergence of the co-vector $\delta \omega = -\partial_\alpha \omega^\alpha$. 
A straightforward calculation of each term leads to

\begin{equation}\label{EQ-delta-ji-45}
\begin{split}
&\delta (j^5i_5+j^4i_4)d\phi=\\
&\frac{1}{\Delta}\left\{\left(-2(r+1)f+D^2\square_6f+\frac{D^4}{\Delta}F(F^2)\right)F(\phi)\right.\\
&\left. - \left(f\square_6 f - (r+3)F^2 +  \frac{D^2}{\Delta} f F(F^2)\right) r \phi\right.\\
&\left. + D^2 F(F(\phi))\right \}.
\end{split}
\end{equation}
Restricting to $X_f$ gives us
\begin{equation*}
\begin{split}
\delta (j^5i_5+j^4i_4)d\phi=\\
- 2(r+1)\left(F\phi\right)^f  - r \left((\square_6f)^f -  (r+3)F_f^2\right)\phi^f.
\end{split}
\end{equation*}

Finally, the formula for $(\square_6 \phi)^f$ becomes
\begin{align}\label{EQ-Mainequation}
\left(\square_{6}\phi\right)^f = \square_{f}\phi^f \nonumber\\
 + 2(r+1)\left(F\phi\right)^f  + r \left((\square_6f)^f -  (r+3)F_f^2\right)\phi^f.
\end{align}
This is the main formula, announced before, linking explicitly the field equations of $\setR^6$ and $X_f$. 
As a byproduct it leads to  an expression relating $f$ and the Ricci scalar of $X_f$, which is derived in the next section.

\subsection{A relation between $R^f$ and $f$}\label{SEC-curvature}
The expression (\ref{EQ-Mainequation}) applies in particular to the massless conformally coupled scalar field for which $r=-1$. With this
particular value it becomes
\begin{equation*}
\left(\square_{6}\phi\right)^f = \square_{f}\phi^f  - \left((\square_6f)^f - 2 F_f^2\right)\phi^f.
\end{equation*}
Now, the Klein-Gordon operator acting on the conformal scalar field, on a (locally conformally flat) manifold $X_f$, is well 
known and reads
\begin{equation*}
\left(\square_{f}  - \frac{R^f}{6}\right)\phi^f,
\end{equation*}
where $R^f$ is the Ricci scalar of $X_f$. Apart from its homogeneity, $f$ is otherwise an arbitrary function, thus one must have
\begin{equation}\label{EQ-R from f}
\frac{R^f}{6} = (\square_6f)^f - 2 F_f^2.
\end{equation}
Since $F:=\widetilde{df}$, this relation allows us to compute the Ricci scalar $R^f$ from the function $f$ only.

Finally, we remark that for the conformal scalar field, the constraint $F$ disappears of the Eq. (\ref{EQ-Mainequation}) due 
to the degree of homogeneity, $-1$,  of the field. This is due to the fact that the conformal scalar field evolves, indeed, on the 
whole null cone $\mathcal{C}$.

\section{The Klein-Gordon form of $(\square_6\phi)^f$}\label{SEC-KGForm}

In the general expression (\ref{EQ-Mainequation}), neither the choice of the manifold defined by $f$, nor the constraints on the field, 
the  action of $F$ and the degree of homogeneity (action of $D$), have been specified. In this section we examine, in the non-conformal 
case ($r\ne -1$), how constraints can be set 
in order to obtain on $X_f$ the Klein-Gordon operator acting on the scalar field $\phi^f$: 
\begin{equation*}
(\square_{6}\phi)^f = (\square_{f} - \xi R^f  + m^2)\phi^f.
\end{equation*}
The result is that, in order to solve the Klein-Gordon equation on a spacetime $X_f$ the homogeneous field $\phi$ solution of $\square_6 \phi = 0$
has to satisfy the constraint 
\begin{equation}\label{EQ-CnstraintFLRW}
F\phi = G(y)\phi,
\end{equation}
where $G(y)$ is a function, homogeneous of degree $-1$, determined by $f$.  
For a FLRW spacetime, the vector $F$ and the function $G$ are explicitly determined in function of the scale factor $a$. An 
explicit realization in a convenient global coordinate system of this constraint is given in Appendix \ref{App-ConstrainFLRW}.
In the following, we first consider the general FLRW spacetimes, then we re-examine the (A)dS spacetimes for 
which a specific and complete treatment is possible.

\subsection{Robertson-Walker spaces}\label{SEC-RW}

When $f$ defines a space which is not maximally symmetric, the terms $(\square_6 f)^f$ and $F^2_f$, which appear  
in the last term of (\ref{EQ-Mainequation}), are not constant. This reflects from (\ref{EQ-R from f}) the non-constancy of the scalar curvature
$R^f$. As a consequence,  the massive Klein-Gordon operator:  
$(\square_{6}\phi)^f = (\square_{f} - \xi R^f  + m^2)\phi^f $, could be recovered only for a proper choice of
the constraint $F\phi$. One can verify that unfortunately the simplest choice: $F\phi=0$, doesn't work, for instance, 
for a FLRW spacetime with scale factor $a(t) = t^\alpha$, which is an approximation of common use in cosmology 
(see Appendix \ref{App-ConstrainFLRW}). Indeed, the choice $F\phi = 0$ can only be made for the (A)dS spaces as we will see in 
Section \ref{SEC-AdS}.  

The choice of the constraint $F\phi$ is indicated by the Eq. (\ref{EQ-Mainequation}) which can be recast under the form 
of the Klein-Gordon equation $ (\square_{f} - \xi R^f  + m^2)\phi^f =0 $ if
\begin{equation}\label{EQ-Condition-KG-FLRW}
\begin{split}
 2(r+1)\left(F\phi\right)^f +  r \left((\square_6f)^f -  (r+3)F_f^2\right)\phi^f =\\
 (-\xi R^f + m^2) \phi^f.
\end{split}
\end{equation}
Then, $F\phi$ must be set in order to have $m^2$ constant.  The simplest choice is
\begin{equation}\label{EQ-Constaint-Fphi=Gphi}
F\phi = G(y)\phi,
\end{equation}
where $G$ is a function homogeneous of degree $-1$ satisfying
\begin{equation*}
\begin{split}
G^f =  \frac{1}{2(r+1) }\left\{ m^2- \left( (r + 6 \xi) (\square_6f)^f \right. \right.\\
 + \left.\left. \left(r(r+3) + 12 \xi\right) F^2_f \right)\right\}, 
\end{split}
\end{equation*}
where we have introduced the coupling constant $\xi$ to the curvature $R^f$ and used Eq. (\ref{EQ-R from f}).
Apart from the case of maximally symmetric spaces considered below, no condition is present to determine 
the function $G$. We then 
resort to a simplicity criterion to choose that function: using the 
Eq. (\ref{EQ-delta-ji-45}) restricted to the cone ($D^2 = 0$), one can set
\begin{equation}\label{EQ-G-FLRW}
G = \alpha \square_6 f + \beta \frac{F^2}{f} + G_0 \frac{1}{f},
\end{equation}
with
\begin{equation*}
\alpha := \frac{(r+6\xi)}{2(r+1)},~ \beta := \frac{r((r+3) + 12\xi)}{2(r+1)},~G_0:=\frac{m^2}{2(r+1)}.
\end{equation*}

The constraint (\ref{EQ-Constaint-Fphi=Gphi}) is then completely specified as soon as the function $f$ is known. In addition, as 
shown in Appendix \ref{App-ConstrainFLRW}, given the metric of some FLRW spacetime in some standard form (see \cite{Ibison:2007dv}), 
one can relate the scale factor for that four dimensional space to the function $f$ in $\setR^6$ which realizes it as a space $X_f$. 
Thus,  the constraint  (\ref{EQ-Constaint-Fphi=Gphi}) can be recast in terms of the scale factor of that FLRW space extended to 
a function of $\setR^6$. All possible extensions of the scale factor obtained in that way restrict of course to the same FLRW metric. 
An explicit expression of (\ref{EQ-Constaint-Fphi=Gphi}) is obtained in Appendix \ref{App-ConstrainFLRW} by that procedure using the 
Minkowskian form of the FLRW metric $ds^2 = a^2ds^2_{{\sss M}ink}$ given in \cite{Ibison:2007dv}.

For the maximally symmetric spaces, (A)dS and Minkowski spaces, the situation is somehow different. Indeed, the method we use in this
section, which grounds on the form (\ref{EQ-Constaint-Fphi=Gphi}) of the constraint, is not the most
straightforward (nor probably the most useful for the well-known Minkowski space). It can nevertheless be adapted to those
spaces. Such a modification is done for completeness in Appendix \ref{App-ConstraintMaxSym}. We now turn to
the more direct treatment of (A)dS spaces.   

\subsection{(A)dS spaces}\label{SEC-AdS}

We retrieve here a result obtained in \cite{Huguet:2016szt} using a coordinates based derivation for (\ref{EQ-Mainequation}) specific to 
(A)dS spaces. Let us recall here, for completeness, these results in the present context.

The de Sitter space can be obtained by setting 
\begin{equation}\label{Eq-fdSitter}
f(y) := \frac{1}{2}(1+H^2)y^5 + \frac{1}{2}(1-H^2)y^4, 
\end{equation}
$H$ being related to the Ricci scalar through $R= -12 H^2$. 
The Anti-de~Sitter space is obtained by changing the parameter $H^2$ to $-H^2$ in the above equation. Note that the Minkowski
space is obtained by setting $H=0$. 

By contrast with the case of the FLRW spacetimes with non-constant scalar curvature, the choice $(F\phi)^f = 0$ can be made
in Eq. (\ref{EQ-Mainequation}). Taking the above expression of $f$ into account Eq. (\ref{EQ-Mainequation}) becomes
\begin{equation}\label{EQ-AdSMink}
\left(\square_{6}\phi\right)^f = \square_{f}\phi^f - r(r+3)F^2 \phi^f,
\end{equation}
with $F^2 = F^2_f= H^2, - H^2$ for dS and AdS spaces respectively (note that $F^2 = 0$ corresponds to the Minkowski space). 
The above equation can be seen as the Klein-Gordon equation for all scalar representations of the (A)dS group for a suitable choice of the 
degree of homogeneity $r$.  Precisely,  since on (A)dS space, the Laplace-Beltrami operator $\square_f$,  and the first order 
Casimir operator $\mathcal{Q}_1$,  are related through $\square_f =  - F^2 \mathcal{Q}_1$, all scalar UIRs can be 
retrieved. To this end, starting from Eq. (\ref{EQ-AdSMink}), let us set the parametrization 
\begin{equation*}
   - r(r+3)F^2\phi^f = (m^2 + 12 \xi F^2)\phi^f, 
\end{equation*}
$m$ and $\xi$ being parameters which, in the massive representations, are identified with mass and coupling to the curvature. 
This equation is solved for $r$ and gives
\begin{equation*}
    r = -\frac{3}{2} + 
    \sqrt{\frac{9}{4} - \left(12 \xi +\frac{m^2}{F^2}\right)}.
\end{equation*}
where the positive root is selected, in the AdS case because it gives the right degree of homogeneity to the conformally 
coupled scalar field, and in the dS case because of the symmetry $r\rightarrow -(r+3)$. The value of $12 \xi + m^2/F^2$ can be used in order 
to classify the involved representations of the (A)dS group (see for instance \cite{Huguet:2016szt}). 

{We finally note that for Minkowski space, for which $F^2 =0$, the Eq. (\ref{EQ-Mainequation}) reduces, for $F\phi = 0$, 
to that of the conformal scalar. So that no mass term can be obtained by setting the constraint $F$ to zero.}
\section{Conclusions}\label{SEC-Conclusion}

In the present paper we have explicitly restricted the scalar operator $\square_6$ defined on $\setR^6$ to locally conformally flat 
four dimensional submanifolds of $\setR^6$ obtained as the intersection of hypersurfaces defined through a function $f$, homogeneous 
of degree one, and the null cone $\mathcal{C}$ of $\setR^6$. 

One consequence is that the determination of the solution of the Klein-Gordon equation: 
${\displaystyle (\square_{f} - \xi R^f  + m^2)\phi^f = 0}$, on a FLRW spacetime 
is reduced to the determination of the solution of the equations
\begin{equation*}
\left\{
\begin{aligned}
& \square_6 \phi = 0\\
& F\phi = G(y) \phi,
\end{aligned}
\right.
\end{equation*}
in $\setR^6$, where $\phi$ is homogeneous of an unspecified degree $r$, $F=\sharp df$ is the operator gradient of $f$ (that is 
$F^\alpha = \partial^\alpha f$ in the basis associated with the Cartesian coordinates $\{y^\alpha\}$ of $\setR^6$)  and $G$ a function, both 
explicitly determined by the scale factor of the FLRW spacetime (see Sec. \ref{SEC-RW}).



\appendix

\section{Definitions and formulas}\label{App-DefAndForm}
The space $\setR^6$ is provided with the natural orthogonal coordinates $\{y^\alpha\}$ and the metric  
$\eta~=~\mbox{diag}(+,-,-,-,-,+)$ which is left invariant under the group O$(2,4)$ and its connected subgroup SO$_0(2,4)$. 
Let us remind that our notations regarding differential geometry follow those of Fecko \cite{Fecko:2006}.

For any vector field $A$, the scalar field $A^2=A_\alpha A^\alpha$ is non necessarily positive, 
and we note $|A|=\sqrt{|A^2|}$. With this convention we have $\left(A/|A|\right)^2=\sgn(A^2)$.

When a metric $g(\cdot, \cdot)$ is present, the one-form associated with a vector $\xi$ is given by $\flat_g\xi := g(\xi,\cdot)$,
that is in components in an arbitrary basis $\{e_a\}$: $(\flat_g\xi)_a = g_{ab} \xi^b$. The co-vector associated with a one-form 
$\alpha$ is given by $\sharp_g\alpha := g^{-1}(\alpha, \cdot)$   that is in components in an arbitrary basis $\{e_a\}$: 
$(\sharp_g\alpha)^a = g^{ab} \alpha_b$. To save notation, when no confusion is present, we denote by a tilde the co-vector of the one-form $\alpha$,
and the one-form associated with a vector $\xi$, that is~:  $\widetilde\alpha\nlb:=\nlb\sharp_g\alpha$,   $\widetilde\xi:=\flat_g\xi$.

The interior product with a vector $\xi$ is denoted by $i_\xi$ and the exterior product with a one-form $\alpha$
by $j_\alpha$. For a vector $\xi$,  $j_\xi := j_{\widetilde\xi}$. 
We note  that the operators $i$ and $j$ are nilpotent.

The Laplace-de Rham operator is defined as $\square:=- (\delta d + d \delta)$, 
on scalars it reduces to $\square:=-\delta d$. 

We also denote by $\omega_f$ the volume form of the induced metric on $X_f$,  and by $S_\sigma$ the hyper-surface volume element: 
\begin{equation*}
S_\sigma := \frac{1}{(n-1)!} \epsilon_{\sigma\mu_1\ldots\mu_{n-1}} e^{\mu_1}\wedge\ldots\wedge e^{\mu_{n-1}},
\end{equation*}
where $n$ is the dimension of the space.

For convenience we reproduce here the relations used in calculations (mostly of appendix \ref{App-Proof-deld6-deld4}). 
We consider a $n$-dimensional
manifold, we denotes by $g$ a metric, $\omega := \sqrt{\vert g \vert} e^1\wedge\ldots\wedge e^n$  its associated volume form with 
$\{e_{a}\}, a=1,\ldots n$  an arbitrary  positively oriented  basis, and $\ast$ its associated Hodge dual operator. 
For a p-form $\alpha$ we define $\hat\eta$ by $\hat\eta\alpha = (-1)^p\alpha$.
Then, using \cite{Fecko:2006} again
the following relations holds~:
$\ast\ast = \sgn(g)(\hat\eta)^{n+1}$,  $\ast^{-1} = \sgn(g) \ast (\hat\eta)^{n+1}$, $\delta := \ast^{-1} d \ast\hat\eta$, 
$\hat\eta\ast = (-1)^n \ast\hat\eta$, $\ast i = - j \ast\hat\eta$,
$j \ast\nlb= \nlb\ast j \hat\eta$,  $\ast \omega = \sgn(g)$.

\section{Proof of $\square_f \phi^f = -(\delta j^\mu i_\mu d\phi)^f$ }\label{App-Proof-deld6-deld4}
To prove the above equality we compute each member separately. We consider the orthonormal 
basis $\{e^a\}, a = 0\ldots 5$ built in Section \ref{SEC-RestrictioToXf}.
We start by the calculation of $\square_f \phi^f$.
From $\square_f \phi^f = -(\delta_f d_f + d_f \delta_f)\phi^f =$ $ -\delta_f d_f \phi^f$, omitting the super(sub)scripts
$f$ for saving notation, one has successively
\begin{align*}
\delta d \phi &= \delta j^\mu i_\mu d\phi\\
&=\delta j^\mu e_\mu(\phi)\\
&=\delta \eta^{\mu\sigma} j_\sigma e_\mu(\phi)\\
&=\delta  e^\mu e_\mu(\phi)\\
&=\ast^{-1}d\ast \hat\eta  e^\mu e_\mu(\phi)\\
&=\ast d e_\mu(\phi) \ast e^\mu \\
&=\ast d e_\mu(\phi) \eta^{\mu\sigma} S_\sigma \\
&=\ast\left\{e_\nu(e_\mu(\phi))e^\nu\wedge \eta^{\mu\sigma} S_\sigma \right.
\left. +\,e_\mu(\phi) \eta^{\mu\sigma} d S_\sigma \right\}.
\end{align*}
In the first term of the r.h.s. one has
\begin{align*}
e^\nu\wedge S_\sigma &=  \frac{1}{3!} \epsilon_{\sigma\mu_1\mu_2\mu_3} e^\nu\wedge e^{\mu_1}\wedge e^{\mu_2}\wedge e^{\mu_3}\\
&= \frac{1}{3!} \left(\frac{1}{4}\epsilon_{\mu\mu_1\mu_2\mu_3} e^\nu\wedge e^{\mu_1}\wedge e^{\mu_2}\wedge e^{\mu_3}\right)\delta^\mu_\sigma\\
&= \omega_f \delta^\nu_\sigma.
\end{align*}
In the second term of the r.h.s
\begin{align*}
d S_\sigma &= d i_{e_\sigma} \omega_f\\
&=\mathcal{L}_{e_\sigma} \omega_f- i_{e_\sigma} d\omega_f\\
&=\mathcal{L}_{e_\sigma} \omega_f.
\end{align*}
Finally, on $X_f$, restoring the super(sub)scripts $f$, one has
\begin{equation*}
\square_f \phi^f = -  \eta^{\mu\nu}\ast\left(e_\nu(e_\mu(\phi^f))\omega_f +  e_\mu(\phi) \mathcal{L}_{e_\nu} \omega_f\right). 
\end{equation*}
Now, a completely analogous calculation in $\setR^6$ leads to 
\begin{equation*}
\delta j^\mu i_\mu d\phi = - \eta^{\mu\nu}\ast\left(e_\nu(e_\mu(\phi))\omega +  e_\mu(\phi)\mathcal{L}_{e_\nu} \omega\right). 
\end{equation*}
All quantities ($d$, $\ast$, $\omega$, \ldots) being now those associated to $\setR^6$. The Lie derivative on the r.h.s. reads
\begin{align*}
\mathcal{L}_{e_\nu} \omega &= \mathcal{L}_{e_\nu} (e^0\wedge\ldots\wedge e^5)\\
&=\left(\mathcal{L}_{e_\nu} \omega_f\right)\wedge e^4\wedge e^5 + \omega_f \wedge  \left(\mathcal{L}_{e_\nu} e^4\wedge e^5\right).
\end{align*}
Using Cartan's formula $\mathcal{L} = id + di$ and owing to the orthonormality of the basis $\{e^a\}$, we have
\begin{equation*}
\left(\mathcal{L}_{e_\nu}  e^4\wedge e^5\right) = i_\nu d (e^4\wedge e^5).
\end{equation*}
To compute the last exterior derivative we re-express the basis one-form using the definition of the basis vectors $\{e_a\}$. For 
$F^2 \ne 0$ the expressions (\ref{EQ-eFe_muFneq0}, \ref{EQ-e4-F2neq0}) lead to 
\begin{align*}
e^4\wedge e^5 &= \frac{1}{\sqrt{\Delta}} \widetilde{D}\wedge\widetilde{F}\\
&=\frac{1}{2\sqrt{\Delta}} dC\wedge df.
\end{align*}
Then, applying the exterior derivative to this expression shows that it vanishes on $X_f$. An analogous calculation leads to the same 
result when $F^2=0$. We are thus left with
\begin{equation*}
\delta j^\mu i_\mu d\phi = -  \eta^{\mu\nu}\ast\left(e_\nu(e_\mu(\phi))\omega \right.
\left. + \, e_\mu(\phi)(\mathcal{L}_{e_\nu} \omega_f)\wedge e^4\wedge e^5\right). 
\end{equation*}
Now, $\mathcal{L}_{e_\nu} \omega_f$ is a 4-form on $X_f$ one has $\mathcal{L}_{e_\nu} \omega_f = a_\nu \omega_f$ with $a_\nu$ 
the proportionality constant. Then, 
\begin{equation*}
\delta j^\mu i_\mu d\phi = - \eta^{\mu\nu}\ast\left(e_\nu(e_\mu(\phi))\omega  + e_\mu(\phi)a_\nu \omega\right). 
\end{equation*}
The same proportionality relation applies in the expression of $\square_f$ which thus reads
\begin{equation*}
\square_f \phi^f = - \eta^{\mu\nu}\ast\left(e_\nu(e_\mu(\phi^f))\omega_f +  e_\mu(\phi) a_\nu \omega_f\right). 
\end{equation*}

Using the relation $\ast_g \omega_g = \sgn(g)$ (were $g$ is the metric) in both equalities and restricting on $X_f$ 
the expression on $\setR^6$, we finally get
\begin{equation*}
\begin{split}
\square_f\phi^f &= -(\delta j^\mu i_\mu d\phi)^f \\ 
&= \left(e^\mu(e_\mu(\phi^f))  + a^\mu e_\mu(\phi^f)\right). 
\end{split}
\end{equation*}

\section{Constraint for FLRW spaces}\label{App-ConstrainFLRW}
Let us remind the reader from \cite{Huguet:2013tv} how 
to obtain a function $f$ which defines a FLRW space as a submanifold of $\setR^6$. We use a result from Ibison \cite{Ibison:2007dv}
which shows explicitly how the metric element of a FLRW spacetime can be cast under the form 
\begin{equation*}
ds^2 = a^2(x^0,\norm{\boldsymbol{x}}) ds^2_{\mathrm{Mink}} ,
\end{equation*}
$\{x^\mu\}$ being a Cartesian coordinate system in Minkowski space. Then, we set the coordinate system to the Minkowskian system defined 
from $\setR^6$ through 
\begin{equation}
x^\mu := \frac{y^\mu}{f_{\sss N}},~~\mathrm{with}~~ f_{\sss N}(y) := \frac{1}{2}(y^5 + y^4).
\end{equation}

The function $f_{\sss N}$ defines
a Minkowski space as a submanifold of $\setR^6$ which can be used as a local chart (see Eq. (\ref{Eq-fdSitter})). 
Note that, when $a$ only depends on $x^0$, then $x^0$ is the conformal time. 
As shown in \cite{Huguet:2013tv} two conformally flat 
spaces defined by some functions $\varphi$ and $h$ in $\setR^6$ are related through a Weyl transformation of the metric which reads
\begin{equation*}
g^\varphi_{\mu\nu} = \left(\frac{h}{\varphi}\right)^2 g^h_{\mu\nu}.
\end{equation*}
Then, one can set 
$f(x) = a(x^0, \norm{\boldsymbol{x}}) f_{\sss N}(x)$ and define $f(y)$ using the definition of the coordinates $\{x^\mu\}$. 
That is
\begin{equation}
f(y):= \frac{f_{\sss N}}{a(\frac{y^0}{f_{\sss N}},  \frac{\norm{\boldsymbol{y}}}{f_{\sss N}}  )}.
\end{equation}
The function $f$ being known, one can calculate all the relevant quantities. 

If we consider for example $a(x^0) = p~(x^0)^q$ we have
\begin{equation}\label{EQ-App-f_FLRW-t^a}
f(y)  = \frac{1}{p} (y^0)^{-q} (y^5 + y^4)^{q+1}.
\end{equation}
Then, if we try to set the constraint $F\phi = 0$ in 
the relation (\ref{EQ-Condition-KG-FLRW}), one obtains
\begin{equation*}
\frac{qr\left(q(r+2) +1\right)+6\xi q(q-1)}{(x^0)^2} = m^2,
\end{equation*}
which makes apparent that the l.h.s.  cannot be a non-null constant $m^2$, as claimed in Section \ref{SEC-RW}.

Returning to the general case, it is convenient for practical purposes to use
the  system of coordinates defined by
\begin{equation}\label{EQ-coordNord}
\left \{
 \begin{aligned}
  X^{\pm} &= \frac{1}{2}(y^4 \pm y^5)\\
  X^\mu &= y^\mu .
 \end{aligned}
\right .
\end{equation} 
In these coordinates $a \equiv a(X^0/X^+, \norm{\boldsymbol{X}}/X^+)$ the non-zero components of $F$ read
\begin{equation*}
\left \{
 \begin{aligned}
  F_\mu &= -\frac{X^+}{a}\partial_\mu \log a\\
  F_+ &= \frac{1}{a} (1 - X^+\partial_+ \log a).
 \end{aligned}
\right .
\end{equation*} 
Here $\partial_b \log a$, with $b= \mu,-,+$ is a shorthand for $\partial_b a/ a$.  
Taking into account that $\square_6 = \partial^\mu\partial_\mu - 2 \partial_+\partial_-$ in these coordinates one can obtain $\square_6 f = 
\partial_\mu F^\mu$ and the 
function $G$ appearing in constraint (\ref{EQ-Constaint-Fphi=Gphi}) for any non-maximally symmetric spacetime.
Then the constraint equation (\ref{EQ-Constaint-Fphi=Gphi}) explicitly reads
\begin{equation*}
\begin{split}
&-\frac{1}{a}\left[X^+a\eta^{\mu\nu}(\partial_\nu \log a)\partial_\mu +  (1 - X^+\partial_+ \log a)\partial_-\right]\phi =\\
&\left[\left((\alpha + \beta) (\partial_\mu \log a)^2 -\alpha  \partial^\mu\partial_\mu\log a\right) \frac{X^+}{a} + G_0 \frac{a}{X^+}\right]\phi,
\end{split}
\end{equation*}
where $\alpha, \beta$ and $G_0$ are the constants defined in Eq. (\ref{EQ-G-FLRW}).

Although probably of little practical interest (because already considered in a more straightforward way in Section \ref{SEC-AdS}),
maximally symmetric cases can be taken into account, as in Appendix \ref{App-ConstraintMaxSym}, by replacing $G$ by
$G_{\sss MS}$. 

\section{Constraint for the maximally symmetric spaces}\label{App-ConstraintMaxSym}

As FLRW spacetimes, the considerations of Section \ref{SEC-RW} should
apply to maximally symmetric spacetimes. Indeed, from the definition of $f$ for these spaces (\ref{Eq-fdSitter}), the commutator 
$[\square_6, F]$ vanishes identically. 
As a consequence, the function $G$ appearing in (\ref{EQ-Constaint-Fphi=Gphi}) must satisfy the compatibility condition
$\square_6 \left(G\phi\right) = 0$.  That is, on the set of homogeneous solutions of $\square_6\phi = 0$, $G$ must satisfy
\begin{equation}\label{EQ-EQforG(y)}
\left((\square_6 G) + 2 \widetilde{dG}\right)\phi = 0,
\end{equation}
which is not the case for the function given in Eq. (\ref{EQ-G-FLRW}). 
In fact,
it is sufficient to obtain a solution for (\ref{EQ-EQforG(y)}) in a neighborhood of $X_f$.  To this end, 
using $\{f, D^2/f^2\}$ as an alternate parametrization of the sets (neighborhoods)
$M_{\lambda,q}$ defined in (\ref{EQ-Def-Mpq}), one can 
extend the function $G$ of Eq.  (\ref{EQ-G-FLRW}) in  
\begin{equation}\label{EQ-Def-G-VoisinageFLRW}
G_{\sss MS} := \left(1 + a_{\sss F} \left(\frac{D^2}{f^2}\right) \right) G =  \left(1 + a_{\sss F} 
\left(\frac{D^2}{f^2}\right) \right)\frac{G_{\sss F}}{f},
\end{equation}
where $G_{\sss F} := G^f$, for $f$ given by (\ref{Eq-fdSitter}), with $H^2$ replaced by the constant $F^2 = - H^2, H^2, 0$ for 
de~Sitter, Anti-de~Sitter, and Minkowski spaces respectively.  

The constant $a_{\sss F}$ is determined by solving (\ref{EQ-EQforG(y)}) at lowest order in  
$(1/f, D^2/f^2)$, it reads 
\begin{equation*}
a_{\sss F} = \frac{G_{\sss F}-F^2}{2r}.
\end{equation*}
Finally, one can check that for the maximally symmetric spaces, the explicit expression (\ref{Eq-fdSitter}), together with  
(\ref{EQ-Mainequation}) and the constraint $F\phi = G_{\sss MS} \phi$, leads to the Klein-Gordon
equation with $R=-12F^2$.

\section*{acknowledgments}
A.B. wish to thank the \textit{Université de Corse} and the \textit{Collectivité Territoriale de Corse} 
for their financial support as well as the \textit{Laboratoire APC-Astroparticule et Cosmologie} for its hospitality.


\bibliography{bibliography}

\begin{thebibliography}{13}%
\makeatletter
\providecommand \@ifxundefined [1]{%
 \@ifx{#1\undefined}
}%
\providecommand \@ifnum [1]{%
 \ifnum #1\expandafter \@firstoftwo
 \else \expandafter \@secondoftwo
 \fi
}%
\providecommand \@ifx [1]{%
 \ifx #1\expandafter \@firstoftwo
 \else \expandafter \@secondoftwo
 \fi
}%
\providecommand \natexlab [1]{#1}%
\providecommand \enquote  [1]{``#1''}%
\providecommand \bibnamefont  [1]{#1}%
\providecommand \bibfnamefont [1]{#1}%
\providecommand \citenamefont [1]{#1}%
\providecommand \href@noop [0]{\@secondoftwo}%
\providecommand \href [0]{\begingroup \@sanitize@url \@href}%
\providecommand \@href[1]{\@@startlink{#1}\@@href}%
\providecommand \@@href[1]{\endgroup#1\@@endlink}%
\providecommand \@sanitize@url [0]{\catcode `\\12\catcode `\$12\catcode
  `\&12\catcode `\#12\catcode `\^12\catcode `\_12\catcode `\%12\relax}%
\providecommand \@@startlink[1]{}%
\providecommand \@@endlink[0]{}%
\providecommand \url  [0]{\begingroup\@sanitize@url \@url }%
\providecommand \@url [1]{\endgroup\@href {#1}{\urlprefix }}%
\providecommand \urlprefix  [0]{URL }%
\providecommand \Eprint [0]{\href }%
\providecommand \doibase [0]{http://dx.doi.org/}%
\providecommand \selectlanguage [0]{\@gobble}%
\providecommand \bibinfo  [0]{\@secondoftwo}%
\providecommand \bibfield  [0]{\@secondoftwo}%
\providecommand \translation [1]{[#1]}%
\providecommand \BibitemOpen [0]{}%
\providecommand \bibitemStop [0]{}%
\providecommand \bibitemNoStop [0]{.\EOS\space}%
\providecommand \EOS [0]{\spacefactor3000\relax}%
\providecommand \BibitemShut  [1]{\csname bibitem#1\endcsname}%
\let\auto@bib@innerbib\@empty
\bibitem [{\citenamefont {Huguet}\ \emph {et~al.}(2006)\citenamefont {Huguet},
  \citenamefont {Queva},\ and\ \citenamefont {Renaud}}]{Huguet:2006fe}%
  \BibitemOpen
  \bibfield  {author} {\bibinfo {author} {\bibfnamefont {E.}~\bibnamefont
  {Huguet}}, \bibinfo {author} {\bibfnamefont {J.}~\bibnamefont {Queva}}, \
  and\ \bibinfo {author} {\bibfnamefont {J.}~\bibnamefont {Renaud}},\ }\href
  {\doibase 10.1103/PhysRevD.73.084025} {\bibfield  {journal} {\bibinfo
  {journal} {Phys. Rev. D}\ }\textbf {\bibinfo {volume} {73}},\ \bibinfo
  {pages} {084025} (\bibinfo {year} {2006})},\ \Eprint
  {http://arxiv.org/abs/gr-qc/0603031} {arXiv:gr-qc/0603031 [gr-qc]}
  \BibitemShut {NoStop}%
\bibitem [{\citenamefont {Huguet}\ and\ \citenamefont
  {Renaud}(2013{\natexlab{a}})}]{Huguet:2013dia}%
  \BibitemOpen
  \bibfield  {author} {\bibinfo {author} {\bibfnamefont {E.}~\bibnamefont
  {Huguet}}\ and\ \bibinfo {author} {\bibfnamefont {J.}~\bibnamefont
  {Renaud}},\ }\href {\doibase 10.1103/PhysRevD.88.124018} {\bibfield
  {journal} {\bibinfo  {journal} {Phys. Rev. D}\ }\textbf {\bibinfo {volume}
  {88}},\ \bibinfo {pages} {124018} (\bibinfo {year} {2013}{\natexlab{a}})},\
  \Eprint {http://arxiv.org/abs/1310.7333} {arXiv:1310.7333 [hep-th]}
  \BibitemShut {NoStop}%
\bibitem [{\citenamefont {Huguet}\ and\ \citenamefont
  {Renaud}(2013{\natexlab{b}})}]{Huguet:2013tv}%
  \BibitemOpen
  \bibfield  {author} {\bibinfo {author} {\bibfnamefont {E.}~\bibnamefont
  {Huguet}}\ and\ \bibinfo {author} {\bibfnamefont {J.}~\bibnamefont
  {Renaud}},\ }\href {\doibase 10.1063/1.4791688} {\bibfield  {journal}
  {\bibinfo  {journal} {J. Math. Phys.}\ }\textbf {\bibinfo {volume} {54}},\
  \bibinfo {pages} {022304} (\bibinfo {year} {2013}{\natexlab{b}})},\ \Eprint
  {http://arxiv.org/abs/1301.7646} {arXiv:1301.7646 [gr-qc]} \BibitemShut
  {NoStop}%
\bibitem [{\citenamefont {Ade}\ \emph {et~al.}(2016)\citenamefont {Ade} \emph
  {et~al.}}]{Planck:2016}%
  \BibitemOpen
  \bibfield  {author} {\bibinfo {author} {\bibfnamefont {P.~A.~R.}\
  \bibnamefont {Ade}} \emph {et~al.} (\bibinfo {collaboration} {Planck
  Collaboration}),\ }\href {\doibase 10.1051/0004-6361/201525830} {\bibfield
  {journal} {\bibinfo  {journal} {A\&A}\ }\textbf {\bibinfo {volume} {594}},\
  \bibinfo {pages} {A13} (\bibinfo {year} {2016})}\BibitemShut {NoStop}%
\bibitem [{\citenamefont {Solà}\ \emph {et~al.}(2017)\citenamefont {Solà},
  \citenamefont {Gómez-Valent},\ and\ \citenamefont
  {de~Cruz~Pérez}}]{Sola:2016jky}%
  \BibitemOpen
  \bibfield  {author} {\bibinfo {author} {\bibfnamefont {J.}~\bibnamefont
  {Solà}}, \bibinfo {author} {\bibfnamefont {A.}~\bibnamefont
  {Gómez-Valent}}, \ and\ \bibinfo {author} {\bibfnamefont {J.}~\bibnamefont
  {de~Cruz~Pérez}},\ }\href {\doibase 10.3847/1538-4357/836/1/43} {\bibfield
  {journal} {\bibinfo  {journal} {Astrophys. J.}\ }\textbf {\bibinfo {volume}
  {836}},\ \bibinfo {pages} {43} (\bibinfo {year} {2017})},\ \Eprint
  {http://arxiv.org/abs/1602.02103} {arXiv:1602.02103 [astro-ph.CO]}
  \BibitemShut {NoStop}%
\bibitem [{\citenamefont {Bayen}\ and\ \citenamefont
  {Haidari}(1985)}]{Bayen:1984dt}%
  \BibitemOpen
  \bibfield  {author} {\bibinfo {author} {\bibfnamefont {M.~F.~C.}\
  \bibnamefont {Bayen}, \bibfnamefont {F.~Flato}}\ and\ \bibinfo {author}
  {\bibfnamefont {A.}~\bibnamefont {Haidari}},\ }\href {\doibase
  10.1103/PhysRevD.32.2673} {\bibfield  {journal} {\bibinfo  {journal} {Phys.
  Rev.}\ }\textbf {\bibinfo {volume} {D32}},\ \bibinfo {pages} {2673} (\bibinfo
  {year} {1985})}\BibitemShut {NoStop}%
\bibitem [{\citenamefont {Binegar}\ \emph {et~al.}(1983)\citenamefont
  {Binegar}, \citenamefont {Fronsdal},\ and\ \citenamefont
  {Heidenreich}}]{Binegar:1982fv}%
  \BibitemOpen
  \bibfield  {author} {\bibinfo {author} {\bibfnamefont {B.}~\bibnamefont
  {Binegar}}, \bibinfo {author} {\bibfnamefont {C.}~\bibnamefont {Fronsdal}}, \
  and\ \bibinfo {author} {\bibfnamefont {W.}~\bibnamefont {Heidenreich}},\
  }\href {\doibase 10.1063/1.525664} {\bibfield  {journal} {\bibinfo  {journal}
  {J. Math. Phys.}\ }\textbf {\bibinfo {volume} {24}},\ \bibinfo {pages} {2828}
  (\bibinfo {year} {1983})}\BibitemShut {NoStop}%
\bibitem [{\citenamefont {Gulamov}\ and\ \citenamefont
  {Smolyakov}(2012)}]{Gulamov:2011ux}%
  \BibitemOpen
  \bibfield  {author} {\bibinfo {author} {\bibfnamefont {I.~E.}\ \bibnamefont
  {Gulamov}}\ and\ \bibinfo {author} {\bibfnamefont {M.~N.}\ \bibnamefont
  {Smolyakov}},\ }\href {\doibase 10.1007/s10714-011-1301-8} {\bibfield
  {journal} {\bibinfo  {journal} {Gen. Rel. Grav.}\ }\textbf {\bibinfo {volume}
  {44}},\ \bibinfo {pages} {703} (\bibinfo {year} {2012})},\ \Eprint
  {http://arxiv.org/abs/1111.0687} {arXiv:1111.0687 [gr-qc]} \BibitemShut
  {NoStop}%
\bibitem [{\citenamefont {Gazeau}\ \emph {et~al.}(2000)\citenamefont {Gazeau},
  \citenamefont {Renaud},\ and\ \citenamefont {Takook}}]{Gazeau:1999mi}%
  \BibitemOpen
  \bibfield  {author} {\bibinfo {author} {\bibfnamefont {J.~P.}\ \bibnamefont
  {Gazeau}}, \bibinfo {author} {\bibfnamefont {J.}~\bibnamefont {Renaud}}, \
  and\ \bibinfo {author} {\bibfnamefont {M.~V.}\ \bibnamefont {Takook}},\
  }\href {\doibase 10.1088/0264-9381/17/6/307} {\bibfield  {journal} {\bibinfo
  {journal} {Class. Quant. Grav.}\ }\textbf {\bibinfo {volume} {17}},\ \bibinfo
  {pages} {1415} (\bibinfo {year} {2000})},\ \Eprint
  {http://arxiv.org/abs/gr-qc/9904023} {arXiv:gr-qc/9904023 [gr-qc]}
  \BibitemShut {NoStop}%
\bibitem [{\citenamefont {Penedones}(2017)}]{Penedones:2016voo}%
  \BibitemOpen
  \bibfield  {author} {\bibinfo {author} {\bibfnamefont {J.}~\bibnamefont
  {Penedones}},\ }in\ \href {\doibase 10.1142/9789813149441_0002} {\emph
  {\bibinfo {booktitle} {{Proceedings, Theoretical Advanced Study Institute in
  Elementary Particle Physics: New Frontiers in Fields and Strings (TASI 2015):
  Boulder, CO, USA, June 1-26, 2015}}}}\ (\bibinfo  {publisher} {World
  Scientific},\ \bibinfo {year} {2017})\ pp.\ \bibinfo {pages} {75--136},\
  \Eprint {http://arxiv.org/abs/1608.04948} {arXiv:1608.04948 [hep-th]}
  \BibitemShut {NoStop}%
\bibitem [{\citenamefont {Fecko}(2006)}]{Fecko:2006}%
  \BibitemOpen
  \bibfield  {author} {\bibinfo {author} {\bibfnamefont {M.}~\bibnamefont
  {Fecko}},\ }\href@noop {} {\emph {\bibinfo {title} {{Differential Geometry
  and Lie Groups for Physicists}}}}\ (\bibinfo  {publisher} {Cambridge
  University Press},\ \bibinfo {year} {2006})\BibitemShut {NoStop}%
\bibitem [{\citenamefont {Ibison}(2007)}]{Ibison:2007dv}%
  \BibitemOpen
  \bibfield  {author} {\bibinfo {author} {\bibfnamefont {M.}~\bibnamefont
  {Ibison}},\ }\href {\doibase 10.1063/1.2815811} {\bibfield  {journal}
  {\bibinfo  {journal} {J. Math. Phys.}\ }\textbf {\bibinfo {volume} {48}},\
  \bibinfo {pages} {122501} (\bibinfo {year} {2007})},\ \Eprint
  {http://arxiv.org/abs/0704.2788} {arXiv:0704.2788 [gr-qc]} \BibitemShut
  {NoStop}%
\bibitem [{\citenamefont {Huguet}\ \emph {et~al.}(2016)\citenamefont {Huguet},
  \citenamefont {Queva},\ and\ \citenamefont {Renaud}}]{Huguet:2016szt}%
  \BibitemOpen
  \bibfield  {author} {\bibinfo {author} {\bibfnamefont {E.}~\bibnamefont
  {Huguet}}, \bibinfo {author} {\bibfnamefont {J.}~\bibnamefont {Queva}}, \
  and\ \bibinfo {author} {\bibfnamefont {J.}~\bibnamefont {Renaud}},\
  }\href@noop {} {\  (\bibinfo {year} {2016})},\ \Eprint
  {http://arxiv.org/abs/1606.07611} {arXiv:1606.07611 [gr-qc]} \BibitemShut
  {NoStop}%
\end{thebibliography}%

\end{document}